\documentclass{aastex63}
\usepackage{graphics}
\usepackage{pgfplotstable}
\usepackage{float}
\usepackage{dcolumn}% Align table columns on decimal point
\usepackage{bm}% bold math

\let\tablenum\relax
\usepackage{siunitx}
\pgfplotsset{compat=1.16}
\accepted{ApJL}

\shorttitle{Optical luminosity-time correlation}
\graphicspath{{./}{figures/}}
\begin{document}

\title{The Optical Luminosity-Time Correlation for More Than 100 Gamma-Ray Burst Afterglows}

\correspondingauthor{Dainotti, M. G.; first and second authors contributed equally to the paper.}
\email{mdainotti@stanford.edu}

\author{M.G. Dainotti}
\affiliation{Interdisciplinary Theoretical \& Mathematical Science Program, RIKEN (iTHEMS), 2-1 Hirosawa, Wako, Saitama, Japan 351-0198}
\affiliation{Astronomical Observatory, Jagiellonian University, ul. Orla 171, 31-501 Krak{\'o}w, Poland; dainotti@oa.uj.edu.pl}
\affiliation{Space Science Institute, Boulder, Colorado}
\affiliation{SLAC National Accelerator Laboratory, 2575 Sand Hill Road, Menlo Park, CA 94025, USA; dainotti@slac.stanford.edu}

\author{S. Livermore}
\affiliation{Department of Physics and Astronomy, Tufts University, 419 Boston Ave, Medford, MA 02155 USA; samantha.livermore@tufts.edu}

\author{D. A. Kann}
\affiliation{Instituto de Astrof\'isica de Andaluc\'ia (IAA-CSIC), Glorieta de la Astronom\'ia s/n, 18008 Granada, Spain}

\author{L. Li}
\affiliation{ICRANet, Piazza della Repubblica 10, 65122 Pescara, Italy; liang.li@icranet.org}

\author{S. Oates}
\affiliation{School of Physics and Astronomy, University of Birmingham, B15 2TT, UK}

\author{S. Yi}
\affiliation{School of Physics and Physical Engineering, Shandong Provincial Key Laboratory of Laser Polarization and Information Technology, Qufu Normal University, Qufu 273165, China}

\author{B. Zhang}
\affiliation{Department of Physics and Astronomy, University of Nevada, Las Vegas, NV 89154, USA}
%\affiliation{Department of Astronomy \& Astrophysics, Las Vegas University,
%---TO BE UPDATED}

\author{B. Gendre}
\affiliation{OzGrav University of Western Australia, Crawley, 6009, WA, Australia}

\author{B. Cenko}
\affiliation{Astrophysics Science Division, NASA Goddard Space Flight Center, MC 661, Greenbelt, MD 20771, USA}
\affiliation{Joint Space-Science Institute, University of Maryland, College Park, MD 20742, USA}

\author{N. Fraija}
\affiliation{Instituto de Astronomia, Universidad Nacional Autonoma de Mexico, Circuito Exterior, C.U., A. Postal 70-264, 04510 Mexico D.F., Mexico}
%TC:endignore

\begin{abstract}
Gamma-Ray Bursts (GRBs) are fascinating events due to their panchromatic nature. Their afterglow emission is observed from sub-TeV energies to radio wavelengths. We investigate GRBs that present an optical plateau, leveraging on the resemblance with the X-ray plateau shown in many GRB light curves (LCs). We comprehensively analyze all published GRBs with known redshifts and optical plateau observed mostly by the \emph{Neil Gehrels Swift Observatory} (\emph{Swift}). We fit 267 optical LCs and show the existence of the plateau in 102 cases, which is the largest compilation so far of optical plateaus. For 56 \emph{Swift} GRBs with optical and X-ray plateaus, we compare the rest-frame end time at both wavelengths ($T^{*}_{\rm opt}$, $T^{*}_{\rm X}$), and conclude that the plateau is achromatic between $T^{*}_{\rm opt}$ and $T^{*}_{\rm X}$. We also confirm the existence of the two-dimensional relations between $T^{*}_{\rm opt}$ and the optical luminosity at the end of the plateau emission, which resembles the same luminosity-time correlation in X-rays (\citealt{Dainotti2013}). The existence of this optical correlation has been demonstrated for the largest sample of optical plateaus in the literature to date. The squared scatter in this optical correlation is smallest for the subset of the Gold GRBs with a decrease in the scatter equivalent to 52.4\% when compared to the scatter of the entire GRB sample.

%The plane becomes a crucial discriminant corresponding to these environments in terms of the best fitting parameters and dispersions. Most GRBs for which the closure relations are fulfilled with respect to astrophysical environments have an intrinsic scatter $\sigma$ compatible within 1$\sigma$ of that of the ``Gold" GRBs, a subset of long GRBs with relatively flat plateaus. We also find that GRBs satisfying closure relations indicating a fast cooling regime have a lower $\sigma$ than ever previously found in literature.%
\end{abstract}

%In this analysis, we consider short GRBs, short GRBs with extended emission and intrinsically short (GRBs that have the intrinsic duration of the prompt emission $< 2$ s in the rest frame).

\keywords{
GRB}

\section{Introduction} \label{sec:intro}
Gamma-Ray Bursts (GRBs) are the most luminous objects in the Universe, with their luminosities spanning over 8 orders of magnitude. Due to their brightness, we can observe GRBs up to high redshift \citep{Tanvir2009Nature}. Thus, GRBs can be good candidates for use as standard candles because they would extend the Hubble diagram beyond SNe Ia, observed up to $z<2.3$ \citep{Riess2018}. To use GRBs as standard candles, we need to better understand their emission mechanisms. GRBs are traditionally classified as Short (SGRBs) and Long (LGRBs), depending on the prompt emission duration: $T_{90}\leq 2$ s  or $T_{90} \ge 2$ s, respectively\footnote{$T_{90}$ is the time over which a burst emits from $5\%$ to $95\%$ of its total measured counts in the prompt emission.} \citep{Mazets1981,Kouveliotou1993}. LGRBs may originate from the collapse of massive stars (the Collapsar model, \citealt{Woosley1993}), while SGRBs could originate from the merger of two NSs or a NS and a black hole (BH) \citep{AbbottMMA}. To distinguish between these different models, we must classify GRBs according to their phenomenology. The GRB prompt emission is observed in $\gamma$-rays, hard X-rays, and sometimes at optical wavelengths. The afterglow is a long-lasting emission in X-rays, optical, and sometimes radio wavelengths following the prompt emission.

GRB LCs observed by the \emph{Neil Gehrels Swift Observatory} (\emph{Swift}) have more complex features than a simple power-law (PL) decay \citep{Sakamoto2007,Zhang2009}. 
\cite{Sakamoto2007} discovered the existence of a flat part in the X-ray LCs of GRBs, the ``plateau'', which is present soon after the decaying phase of the prompt emission. The \emph{Swift} plateaus generally last from hundreds to a few thousands of seconds \citep{Willingale2007}, hereafter W07, and are followed by a PL decay phase.
Several models have been proposed to explain the plateau, one being the long-lasting energy injection from the central engine by fall-back mass accretion onto a BH. This energy injection will be released into the external shock, where a single relativistic blast wave interacts with the surrounding medium \citep{Zhang2001,Liang2007,Oates2012}. Another possibility is that the energy injection is produced by the spin-down luminosity of a millisecond newborn NS, the so-called magnetar \citep[e.g.,][]{Rowlinson2014,Rea2015,Stratta2018,Fraija2020}. 
In the investigation of the physical mechanisms that drive GRBs, the plateau found at X-ray and optical wavelengths has been highlighted as a feature that could standardize the varied GRB population. \cite{Dainotti2016, Dainotti2017,Dainotti2017C} and \cite{Li2018} explored the relation between the luminosity $L_a$ and rest-frame time $T^{*}_a$ both measured at the end of the plateau (known as the Dainotti relation). We denote the rest-frame time with an asterisk.
\cite{Rowlinson2014} showed that the Dainotti relation in X-rays can be naturally recovered within the magnetar scenario with a slope of $-1$. Within the cosmological context this correlation has already been applied to construct a GRB Hubble diagram out to $z>8$ \citep{cardone2009revised,cardone2010constraining,postnikov2014nonparametric,dainotti2013slope}. 
We investigate this correlation at optical wavelengths to determine how common the plateau is in optical LCs, and how tight the Dainotti relation is for a large optical sample.
This work investigates if a similar correlation in the optical can be determined and can be applied as a reliable cosmological tool in the future.

As determined in \cite{Dainotti2016, Dainotti2017,Dainotti2017C}, it is necessary to select a sub-sample of GRBs with very well-defined characteristics from a morphological and/or a physical point of view to obtain a GRB class that can be standardized, because the tightness of the correlations may also depend on how the sample is divided into classes. %Thus, we divide our sample into classes.
The Long/Short classification has been challenged over the years with the discovery of several sub-classes, that may arise from different progenitors or the same progenitors with different surroundings. Such categories are: SGRBs with extended emission (SEE, \citealt{Norris2006,Levan2007,Norris2010}) with mixed features between SGRBs and LGRBs; Intrinsically Short (IS) GRBs, with $T^*_{90}=T_{90}/(1+z)<2$ s; X-ray flashes (XRFs) with unusually soft spectra and greater fluences in the X-ray band ($2-30$ keV) than in the gamma-ray band ($30-400$ keV, \citealt{Heise2001}); X-Ray Rich GRBs (XRRs) which are intermediate in spectral hardness between XRFs and usual GRBs (\citealt{liu2019grb}); Ultra-Long GRBs (ULGRBs) with a very long prompt duration ($T_{90}>1000$ s, \citealt{Gendre2019}); and GRBs associated with Supernovae, GRB-SNe \citep{Cano2017}. Moreover, there are LGRBs for which an associated SN was not detected, but should have been detected given the observational limits. Examples are the nearby SN-less GRB 060505 and GRB 060614 \citep{Kann2011,ofek2007}; these cases highlight the possibility of LGRBs with and without SNe. The categories of GRB-SNe are: A) strong spectroscopic evidence for an SN associated with the GRB, B) a clear LC bump as well as some spectroscopic evidence suggesting the Long-GRB-SNe association, C) a clear bump in the LC consistent with the GRB-SN associations, but no spectroscopic evidence of the SN, D) a significant bump in the LC, but the properties of the SN are not completely consistent with other GRB-SNe associations or the bump is not well sampled or there is lack of a spectroscopic redshift of the GRB; E) a bump, with low significance or inconsistent with other GRB-SNe identifications, but with the presence of a GRB spectroscopic redshift (\citealt{hjorth2012grb}).

A different classification based on physical mechanisms related to the GRBs' progenitors has been proposed \citep{Zhang2009,Kann2011,Li2020}, according to which GRBs are divided into Type I, powered by compact object mergers: the merger of two NSs or a NS and a BH, and in Type II, characterized by the collapse of massive stars. Type I GRBs include SGRBs, SEE, and IS, while Type II include the LGRBs, GRB-SNe, and XRFs.
A diagram clarifying this classification is shown in Fig. 8 of \cite{Zhang2009}. To homogenize the morphological classification with the one that may arise from different progenitors or the same progenitors with different environments, we ascribe the GRB types in our sample to the Type I or Type II categories.

In \S \ref{sample selection} we detail our sample and data analysis, in \S \ref{methodology} the methodology, in \S \ref{LT correlation} the results of the $L_{\rm opt}-T^{*}_{\rm opt}$ correlation and in \S \ref{conclusion} we summarize our conclusions.

\section{\label{sample selection}Data Analysis and Sample Selection}

%"We considered $\#$ out of the 427 total optical observations of GRBs by \emph{Swift} at this time; the remaining $\#$ GRBs require additional data reduction and may be addressed in a future paper."

We built a comprehensive sample of optical GRB LCs with known redshifts by searching the literature for all GRBs detected between May 1997 to January 2019 by several satellites such as the {\it Swift} Ultra-Violet/Optical Telescope (UVOT), or ground-based telescopes/detectors (e.g., GROND). In our final sample the redshifts of the GRBs span from $z=0.06$ to $z=8.23$ and the LCs employed are found in: \citet[][2021a, 2021b in prep.]{Kann2006,Kann2010,Kann2011}; \cite{Li2012,Li2015,Li2018b}; \cite{Oates2009,Oates2012}; \cite{Zaninoni2013}; and \cite{Si2018}. We then determine the existence of a plateau by fitting the LCs with the phenomenological \footnote{The W07 model makes no assumptions on the underlying physics.} W07 model, see Sec. \ref{methodology}. 

Below, we summarize the data analysis used by \cite{Li2012,Li2015,Li2018b}, \cite{Kann2006,Kann2010,Kann2011}, \cite{Oates2012}, \cite{Zaninoni2013}, and \cite{Si2018}. For GRBs that overlap between these samples, we choose the ones with the greatest coverage, especially in the plateau, and where the $\chi^2$ value for the W07 fitting is the smallest. In some cases, more coverage introduces more scatter which reduces the quality of the fit; in these cases, we select the individual LCs rather than the combined LCs. We include 5 combined LCs in our final sample.

We use 10 GRBs from \citet[][2020 in prep.]{Li2012,Li2015,Li2018b} that meet our requirements defined in Sec. \ref{methodology}. Following \cite{Li2012,Li2015,Li2018b}, we correct for Galactic extinction for the optical and NIR magnitudes, and for host-galaxy extinction correction through an extinction parameter $A_{\rm v}$, assuming $R_{\rm v}=3.1$. The flux contribution coming from the host galaxy at very late times ($\sim 10^{6}$ s after the GRB trigger) for some GRBs has also been subtracted. For the GRBs that were not already corrected for host extinction in the papers cited previously, we computed the extinction factor as $-2.5*Log(A_{\rm v})$ in flux density space. 

We use 57 LCs from \cite{Kann2006,Kann2010,Kann2011}, and Kann et al. 2021a, 2021b (in prep.). Following \cite{Kann2006}, for each afterglow, the multiband LCs are fit with, depending on the detected features, a single PL, a smoothly broken PL, or a series of these. Additionally, if necessary, a constant host-galaxy component is added, and a special supernova-model fit is applied if such a SN is detected following the GRB (see \citealt{Kann2019} for a specific example). The afterglow itself is assumed to evolve achromatically, and therefore the parameters of the afterglow evolution (decay slopes, break time and smoothness) are shared among all bands (host-galaxy and SN parameters are individual to each band). These fits result in a spectral energy distribution (SED) which is determined by the entirety of the data; the SED is assumed to be constant. The SED is then used twofold: first, it allows (after necessary host-and SN-component removal) to shift other bands to the $R_{\rm C}$ band, for which there are essentially always measurements, creating a compound LC with maximised data density and temporal coverage. Furthermore, the SED can be analyzed to determine the line-of-sight extinction in the host galaxy. Then, the LCs are corrected for host-galaxy extinction.

From \cite{Oates2012} we use 3 GRBs which were constructed from multi-filter LCs, following \cite{Oates2009}. The main steps performed are to normalize the multi-filter LCs to the $v$ filter and then to group them using a bin size of $\Delta\rm{t}/\rm{t}=0.2$. The LCs are then normalized to the $R_{\rm C}$ filter relative to the LCs from the \cite{Kann2006,Kann2010,Kann2011} sample which overlaps with the \cite{Oates2012} sample. In \cite{Oates2009}, for each GRB, the onset of the prompt $\gamma$-ray emission (the start time of the $T_{90}$ parameter) is equal to the start time of the UVOT LC. However, here we convert it using the BAT trigger time as the start time of the UVOT LCs to have consistent BAT trigger time, as the other LCs in the sample. To correct for host extinction, for these 3 GRBs we use the same values as \cite{Oates2012}. 

We use 19 GRBs from \cite{Zaninoni2013}. In this paper, optical data is gathered from the literature and from various telescopes, and all units are converted from magnitudes to flux densities; the data are not initially corrected for reddening. SEDs are created at early and late times for each GRB, only using optical filters for which data were available; spectral index values $\beta_{opt}$ are derived from fitting these SEDs, corrected for host and Galactic extinction.

We use 16 LCs investigated in \cite{Si2018}. Their data come from \cite{Li2012} and \cite{Kann2006}. 
%A correction for Galactic extinction is made to the optical data using the reddening maps from \cite{Schlegel1998}, and \cite{Schlafly2011}. 
We corrected this data for host extinction following \cite{Kann2006}.

\section{\label{methodology}Methodology}

\begin{figure*}[!t]
\centering
\includegraphics[width=0.44\textwidth,angle=0,clip]{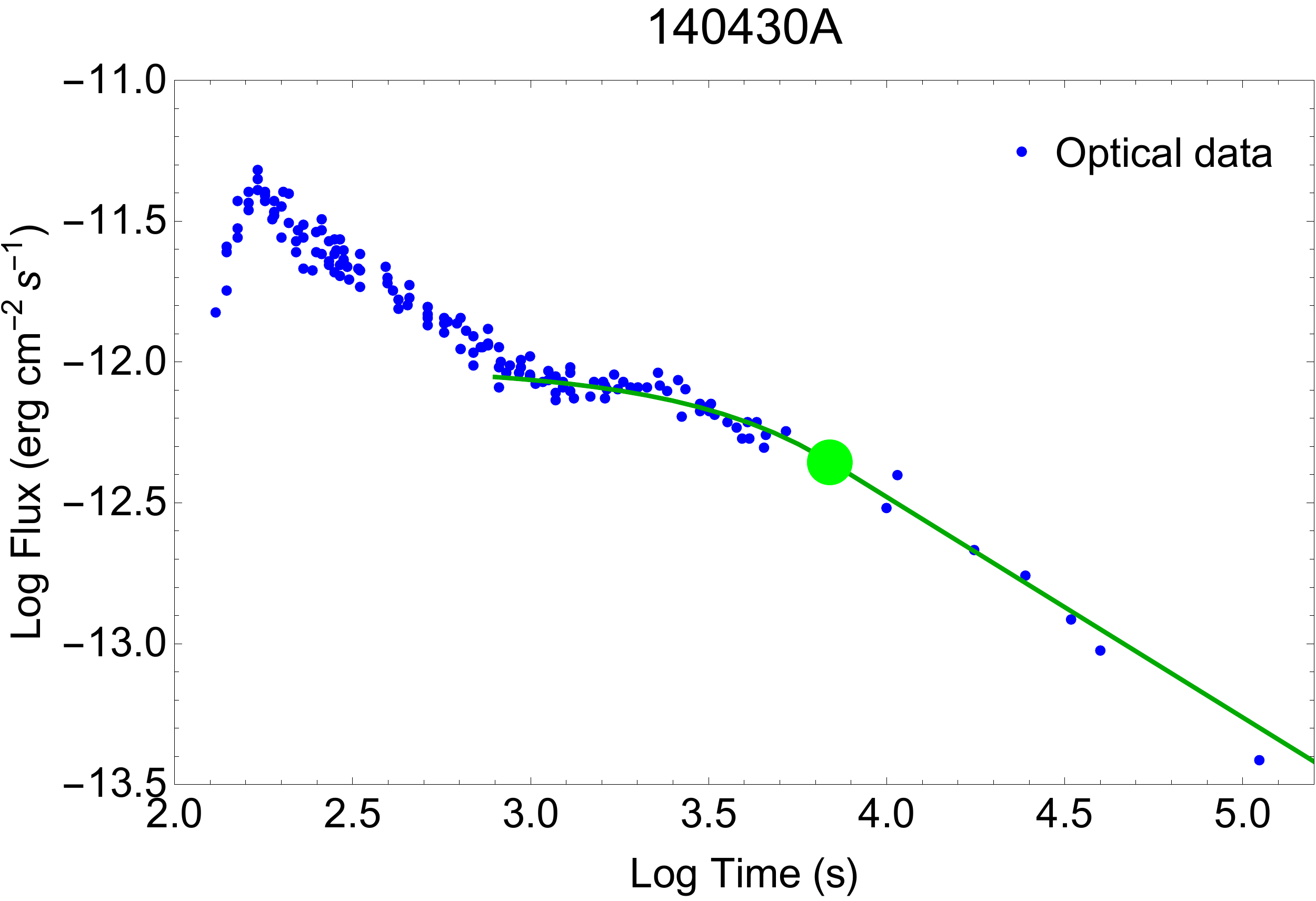}
\includegraphics[width=0.44\textwidth,angle=0,clip]{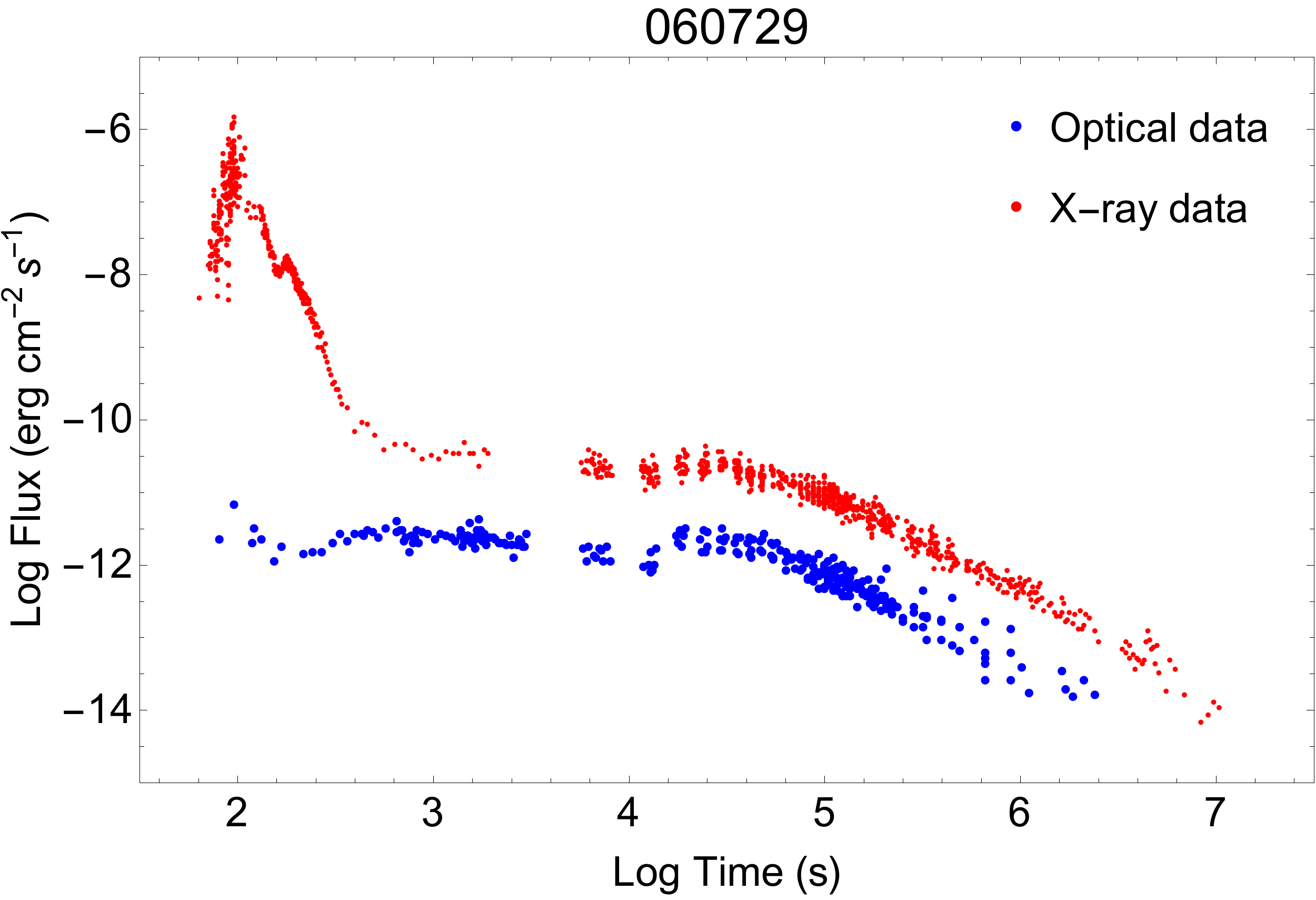}
\includegraphics[width=0.44\textwidth,angle=0,clip]{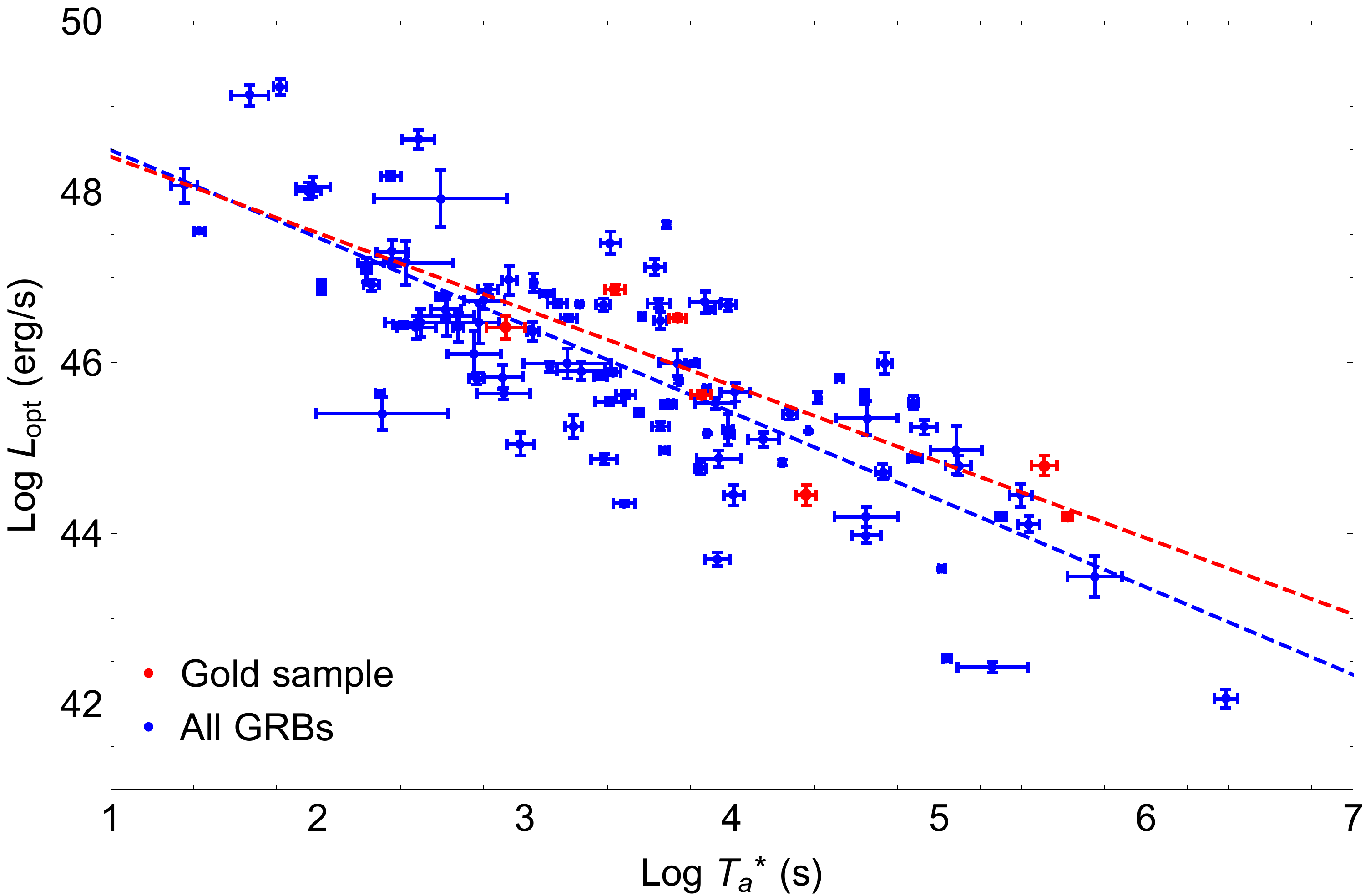}
\includegraphics[width=0.44\textwidth,angle=0,clip]{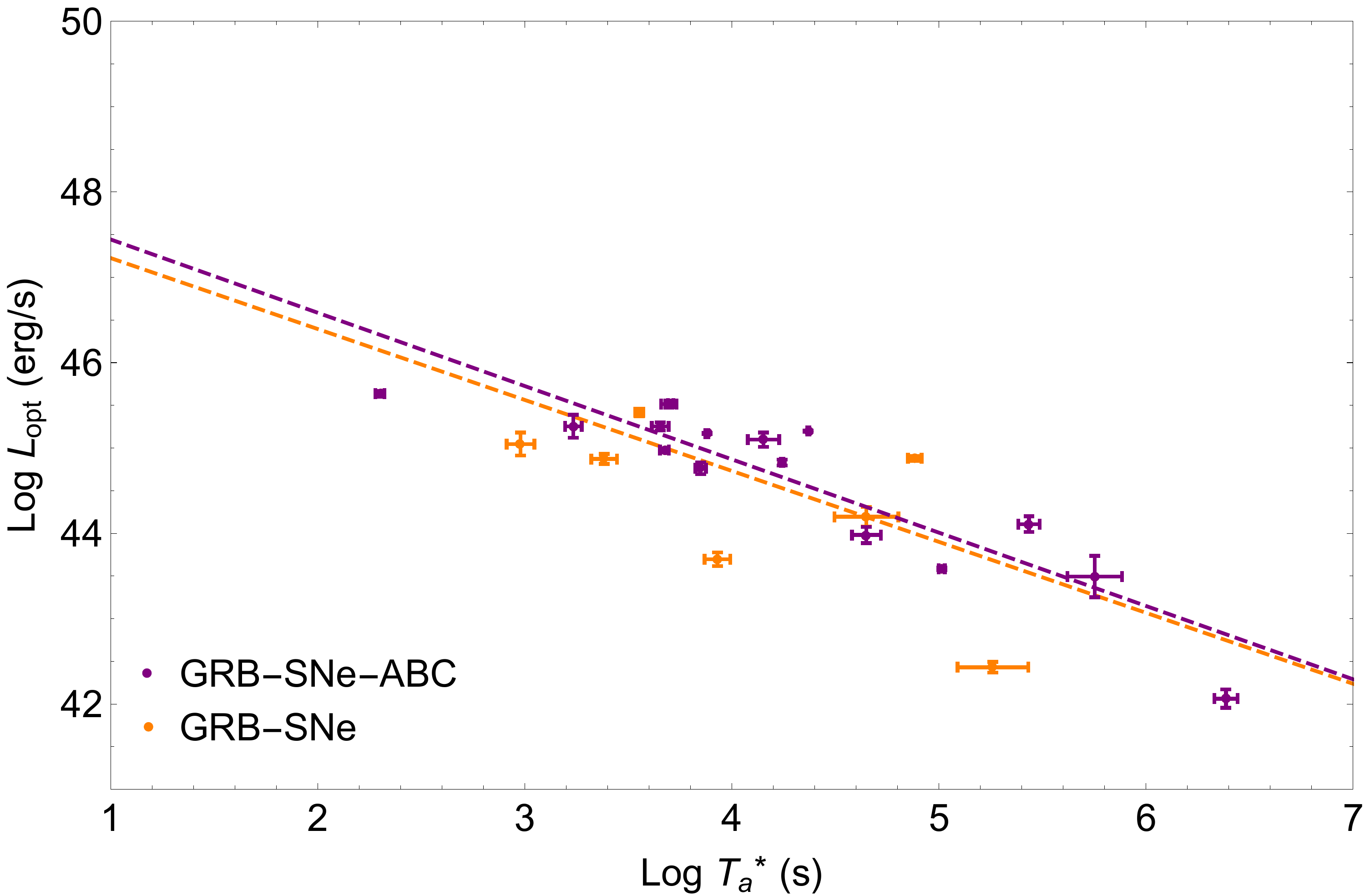}
\caption{{\bf Upper left}: the W07 fit for a well-sampled optical plateau shown as a green line, with the green dot representing  $(T^{*}_{\rm opt},L_{\rm opt})$. Optical data are from Kann et al. (2021a, in prep.). The fits were performed with error bars, which are not shown so as not to clutter the graph. {\bf Upper right}: another example of a well-sampled LC with the coincident observation of X-ray plateau. The optical LC is from \citealt{Zaninoni2013}, while the X-ray LC is from the XRT repository. {\bf Lower panel}: the $L_{\rm opt}- T^{*}_{\rm opt}$ relation for the gold and the total sample (left panel) and for the GRB-SNe total and the GRB-SNe (A,B,C) (right panel). 
The best-fit lines are calculated using a linear model fit in log scale  and plotted in matching colors as dashed lines.}
\label{totalgold}
\end{figure*}

Since the LCs are from different sources in different units, we converted all fluxes into erg cm$^{-2}$ s$^{-1}$ in the R band. We fit the W07 model in the observer frame. Its functional form is:

\begin{equation}
f_{i}(t) = \left \{
\begin{array}{ll}
\displaystyle{F_i \exp{\left ( \alpha_i \left( 1 - \frac{t}{T_i} \right) \right )} \exp{\left (
- \frac{t_i}{t} \right )}} & {\rm for} \ \ t < T_i \\
~ & ~ \\
\displaystyle{F_i \left ( \frac{t}{T_i} \right )^{-\alpha_i}
\exp{\left ( - \frac{t_i}{t} \right )}} & {\rm for} \ \ t \ge T_i. \\
\end{array}
\right .
\label{eq: fc}
\end{equation}
\noindent 
This function $f(t)=f_a(t)+f_p(t)$ is the sum of the two functions that represent both the prompt, $f_p$, and the afterglow emission, $f_a$. We focus on the afterglow. $f(t)$ contains sets of four free parameters $(T_a,F_a,\alpha_a,t_a)$ for each of the two functions $f_a$ and $f_p$, where $T_a$ is the time end of the plateau, $F_a$ is its associated flux, $\alpha_{a}$ is the temporal PL decay index after the plateau, and the time $t_{a}$ is the initial rise timescale of the afterglow. In the majority of cases $t_a$ is compatible with zero, thus it is set as a fixed parameter. The time $T_t$ is the time where $f_p(T_t)=f_a(T_t)$. Its associated flux is $F_t$. We do not fit the LCs with fewer than five data points because this would be too few compared to the fit parameters. Then, we exclude the cases when the fitting procedure fails or the determination of $1 \sigma$ confidence
intervals does not fulfill the $\chi^{2}$ rules; see the XSPEC manual\footnote{http://heasarc.nasa.gov/xanadu/xspec/manual/XspecSpectralFitting.html}. Out of the 267 GRBs analyzed, 102 LCs with well-defined plateaus constitute our final sample, composed of: 35 LGRBs, 9 SGRBs (\citealt{jensen2001afterglow}, \citealt{kaneko2015short}, \citealt{Levan2007}, \citealt{Norris2006}, \citealt{Norris2010},  \citealt{Zhang2009}), 1 SGRBs associated with a Kilonova (\citealt{rossi2020comparison}), 12 XRFs (\citealt{bi2018statistical}, \citealt{Levan2007}, \citealt{ruffini2016classification}), 44 XRRs (\citealt{bi2018statistical}), 23 GRB-SNe (\citealt{Cano2017}, \citealt{hjorth2012grb}, \citealt{klose2019four}), and 4 ULGRBs (\citealt{Gendre2019}, \citealt{gruber2011fermi}). Some GRBs are repeated because they can belong to multiple classes. See Figure \ref{totalgold} for two examples of well-defined plateaus in our sample. We reject 59 LCs for PL behavior, 52 for having too few points or being too scattered, and 54 for having $\Delta \chi^2$ not fulfilling the $\chi^{2}$ prescriptions.

Once we fitted the LCs, we compute from the the optical observed flux $F_{\rm opt}$ (erg cm$^{-2}$ s$^{-1}$) the optical luminosity in the $R_{\rm C}$ filter (one GRBs is in V and another one in H band), $L_{\rm opt}$ (in units of erg s$^{-1}$) using the following:

\begin{equation}
L_{\rm opt}= 4 \pi D_L^2(z) \, F_{\rm opt} (T_{\rm opt}) \cdot \textit{K}
\label{eq: la}
\end{equation}
\noindent at the time $T^{*}_{\rm opt}$ at the end of the optical plateau, where $D_L(z)$ is the luminosity distance, assuming a flat $\Lambda$CDM cosmological model with $\Omega_M=0.3$ and $H_0=70$ km s$^{-1}$ Mpc$^{-1}$.
The k-correction $K$ \citep{bloom2001prompt} is:
\begin{equation}
K = \frac{1}{(1+z)^{1-\beta_{opt}}}
\label{eq: la2}
\end{equation}

\noindent where $\beta_{opt}$ is the optical spectral index of the GRB. The optical spectral parameters are gathered from the literature; for GRBs where $\beta_{opt}$ is unknown, we average values of the whole sample and we use the mean square error (MSE) as the error: $\beta_{opt}=0.79\pm0.03$. \\

The Gold Sample is a sub-sample of GRB LCs with at least four points at the start time of the plateau emission and with plateau inclination $\leq 41^{\circ}$ (for details, see \citealt{Dainotti2016}). The inclination is defined using trigonometry as $\frac{\Delta F}{\Delta t} = \frac{F_{t}-F_{a}}{T_{a}-T_{t}}$. These criteria ensure the plateau is well-defined and shallow enough not to be considered a simple PL. The Gold Sample consists of 7 GRBs. 

\noindent 

\section{\label{LT correlation} The Luminosity-Time Correlation for Optical Plateaus}
Following \cite{Dainotti2017} we investigate the PL relation between the optical luminosity and rest-frame time at the end of the optical plateau : the $L_{\rm opt}-T^{*}_{\rm opt}$ correlation for 102 GRBs, see
Figure \ref{totalgold}. The best-fit parameters are calculated using the linear least square method with the command LinearModelFit in Mathematica 12.1 using the variables in the log scale for convenience. 
LinearModelFit constructs a linear model of the form $<y>=\gamma_0+\gamma_1$ $x_1+\gamma_2$ $x_2+...$ that fits the $y_i$ for successive x values 1, 2... under the assumption that the original $y_i$ are independent normally distributed. In our case $y_i=\log L_{i,opt}$ and $x_i=\log T^{*}_{i,opt}$, where $i$ denotes the GRBs in the sample.
In this paper uncertainties are quoted at $1\sigma$ and we do not account for selection biases and redshift evolution as discussed in \cite{Dainotti2013,Dainotti2017C}. We will address this problem in a forthcoming paper. Here we investigate whether the luminosity-time correlation holds for a large sample of optical plateaus, if there are classes favored because they have small squared scatter, hereafter $\sigma^{2}$, and what the similarities and differences between the luminosity-time correlation in X-rays and in optical are.

%\begin{figure}
%\centering
%\includegraphics[width=0.32\textwidth,angle=0,clip]{oct12_SNeABC_prob.png}
%\includegraphics[width=0.32\textwidth,angle=0,clip]{oct12_gold_prob.png}\\
%\includegraphics[width=0.32\textwidth,angle=0,clip]{oct12_SNe_prob.png}
%\includegraphics[width=0.32\textwidth,angle=0,clip]{oct12_xrf_prob.png}
%\includegraphics[width=0.32\textwidth,angle=0,clip]{oct12_xrr_prob.png}
%\includegraphics[width=0.32\textwidth,angle=0,clip]{oct12_long_prob.png}
%\includegraphics[width=0.32\textwidth,angle=0,clip]{oct12_total_prob.png}
%\includegraphics[width=0.32\textwidth,angle=0,clip]{oct12_short_prob.png}
%\caption{Plots of the likelihood contours and probability distributions are shown in dark and light blue corresponding to 68\% and 95\% confidence intervals, respectively, shown in decreasing order of $\sigma_{\rm int}$ for the SNe classes A, B, and C and the Gold sample (upper panel), the SNe total, the XRFs and XRR (middle panel) and LGRBs, the total sample and SGRBs (lower panel). The parameters, $a$, $c$, and $sv=\sigma_{\rm int}$ are the slope of the correlation, the normalization and the intrinsic scatter, respectively.}
%\label{contours}
%\end{figure}

\begin{figure}
\centering
\includegraphics[width=0.45\textwidth,angle=0,clip]{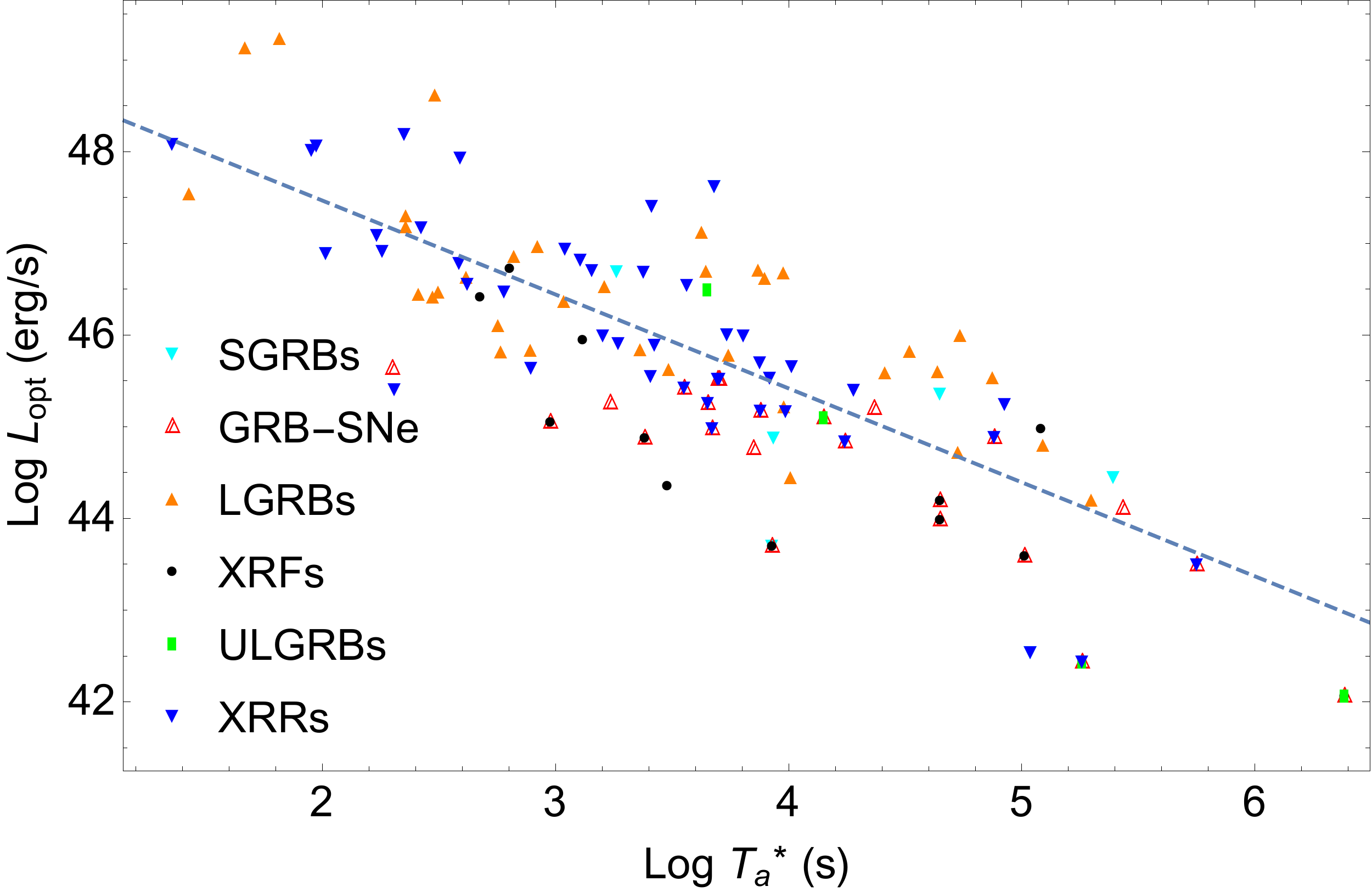}
\includegraphics[width=0.45\textwidth,angle=0,clip]{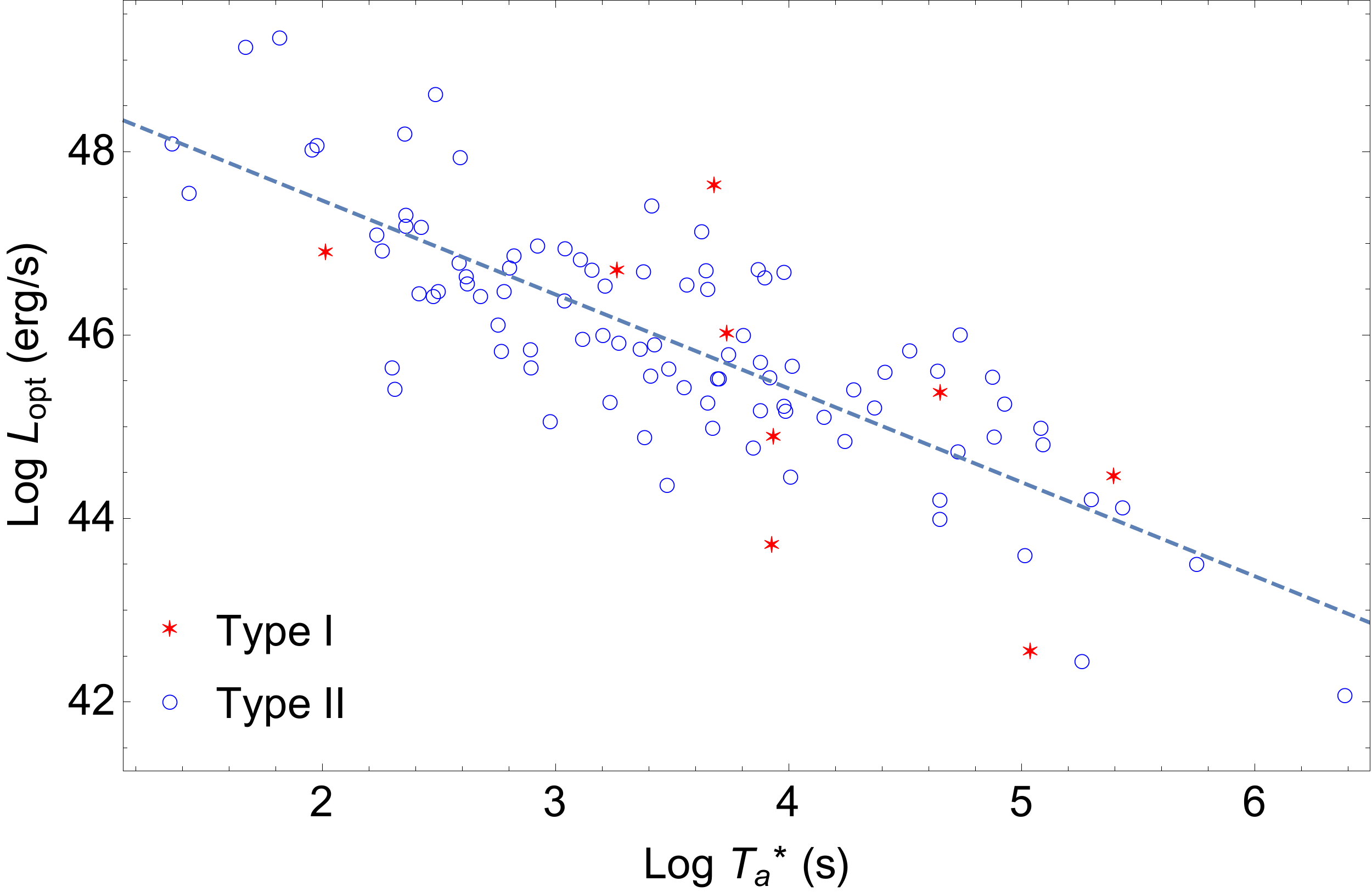}
\includegraphics[width=0.45\textwidth,angle=0,clip]{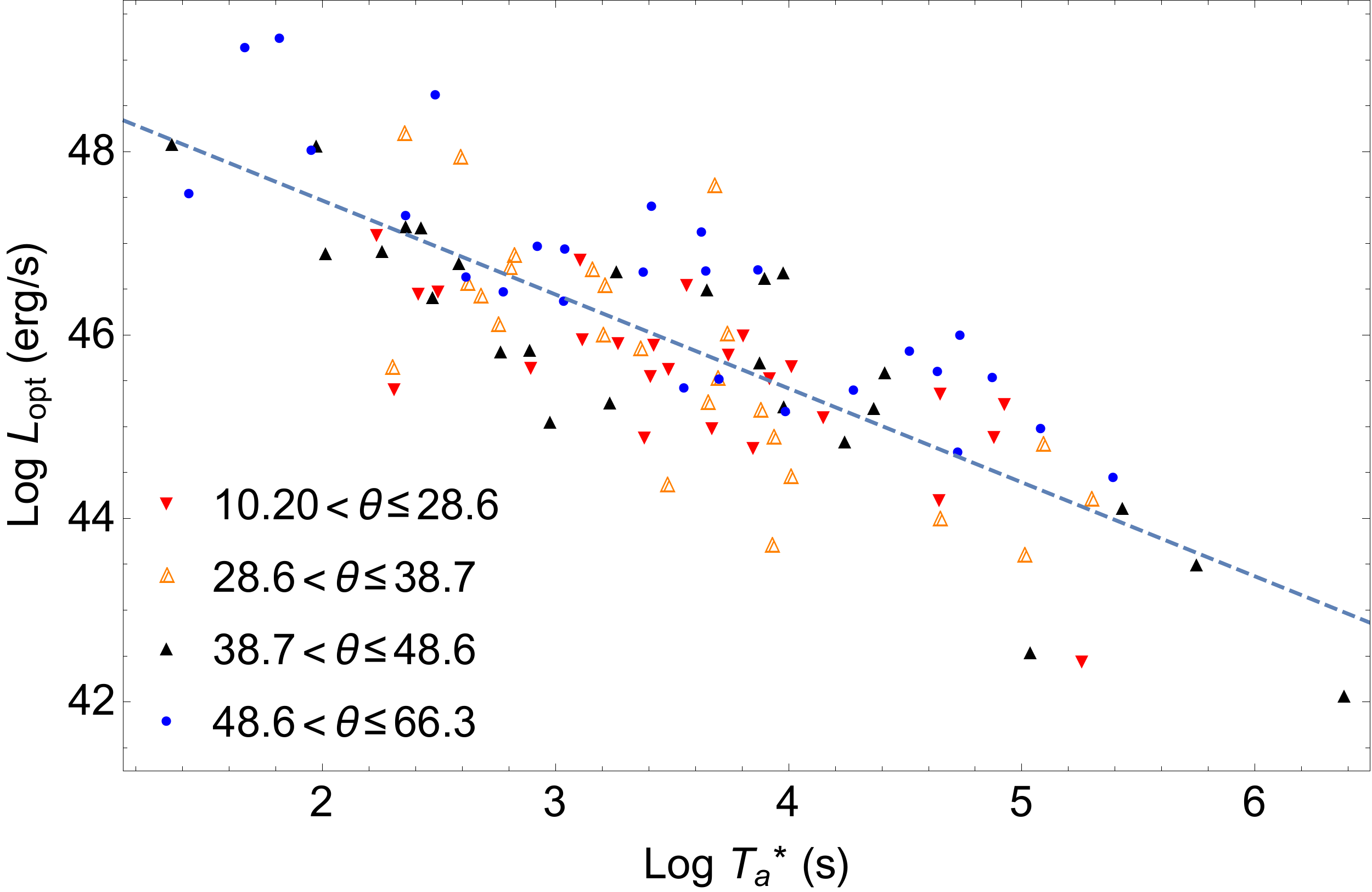}
\includegraphics[width=0.45\textwidth,angle=0,clip]{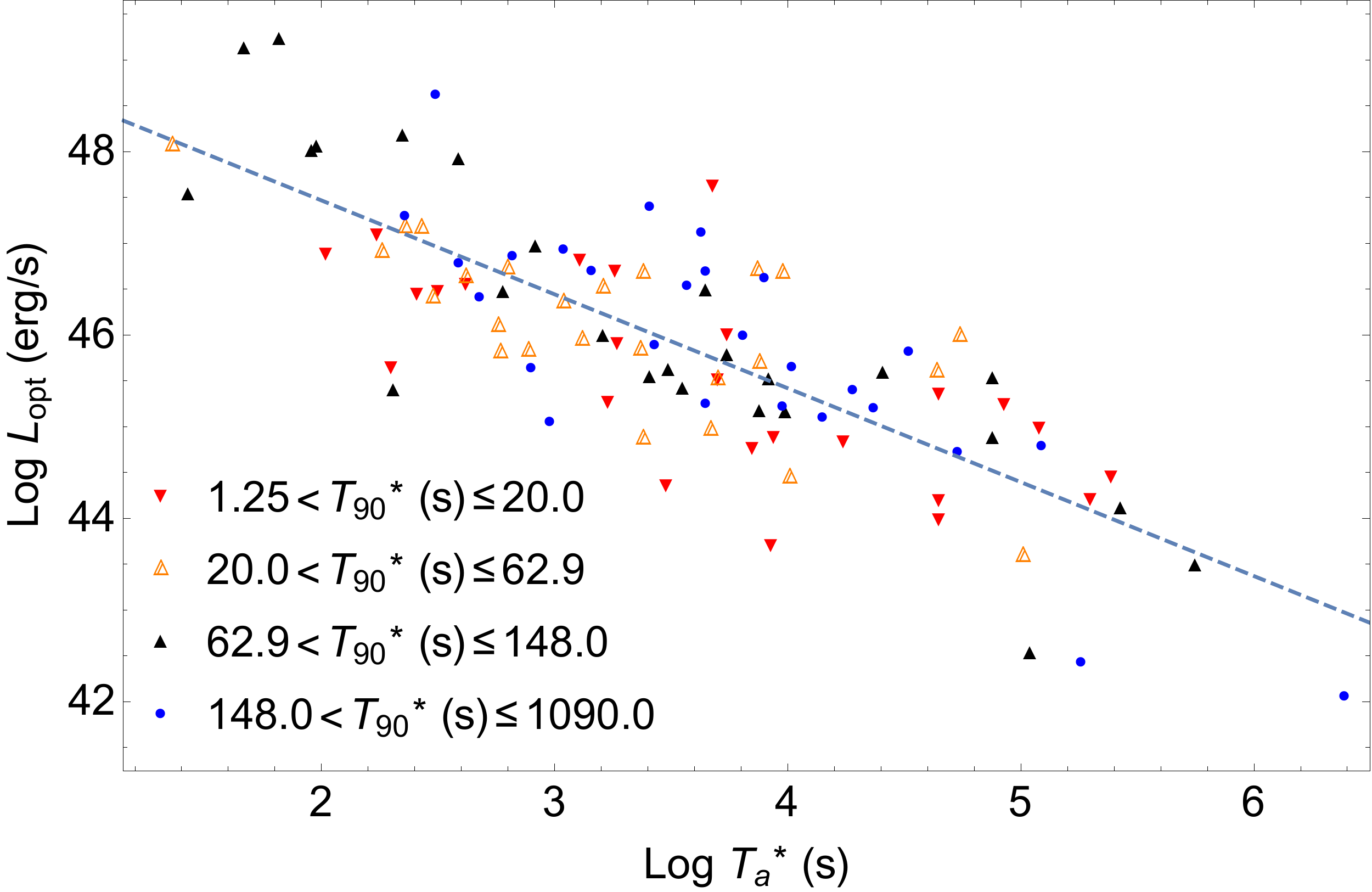}
\caption{The $L_{\rm opt}- T^{*}_{\rm opt}$ relation, plotted according to class, type (I or II), plateau angle, and $T^{*}_{90}$. The best-fit lines shown here represent the slope of the total sample of 102 GRBs, calculated using the a linear model fit in log scale in Mathematica 12.1, and are plotted as dashed lines. %\textcolor{purple}{\bf ALEX: I'm somewhat confused here. you show splits into different ``variants'' but only one fit line each. Shouldn't there be more similar to Fig. 1 bottom?}{\color{red}SamL: the trendline in these plots is the trendline of the total sample of All GRBs}
}
\label{classanglet90}
\end{figure}
\noindent

\begin{table}
\begin{tabular}{| l | l | l | l | l | l | l| l |l | l} \hline
Class & {$a_{opt}$} & $C_o$ & {\it N} & $\rho$ & {\it P} & $|\Delta\rho|$ & $\sigma^2$ of the fit & $\delta{\sigma^2}$ \\ 
\hline
{All GRBs} & $-1.02\pm0.16$ & $49.52\pm0.58$ & 102 & $-0.77$ & $2.7\times10^{-23}$ & 0 & 0.63 & $0.0\%$ \\ \hline
{Gold} & $-0.89\pm0.64$ & $49.31\pm2.75$ & 7 & $-0.86$ & $1.6\times10^{-2}$ & $10.5\%$ & 0.30 & $-52.4\%$ \\ \hline
{SGRBs} & $-1.11\pm1.06$ & $49.73\pm4.34$ & 9 & $-0.80$ & $4.4\times10^{-2}$  & $3.75\%$ & 0.49 & $-22.2\%$\\ \hline
{LGRBs} & $-0.86\pm0.26$ & $49.25\pm0.91$ & 35 & $-0.75$ & $1.1\times10^{-7}$ &  $2.7\%$ & 0.86 & $+36.0\%$ \\ \hline
{XRFs} & $-0.97\pm0.43$ & $48.53\pm1.81$ & 12 & $-0.82$ & $5.7 \times 10^{-4}$ & $6.09\%$ & 0.76 & $+20.6\%$ \\ \hline
{GRB-XRR} & $-1.14\pm0.24$ & $49.89\pm0.84$ & 44 & $-0.80$ & $4.0\times10^{-12}$ & $3.8\%$ & 0.81 & $+28.6\%$ \\ \hline
{GRB-SNe} & $-0.83\pm0.27$ & $48.06\pm1.16$ & 23 & $-0.77$ & $2.3\times10^{-6}$ & $0\%$ & 1.00 & $+58.7\%$ \\ \hline
{GRB-SNe-ABC} & $-0.86\pm0.24$ & $48.30\pm1.04$ & 16 & $-0.85$ & $2.0\times10^{-6}$ & $9.4\%$ & 0.79 & $+25.4\%$ \\ \hline
\end{tabular}
\caption{The best-fit parameters for each sub-sample are calculated using a linear model fit in log scale in Mathematica 12.1: $a_{\rm opt}$ is the slope of the correlation, $C_0$ is the normalization constant, $N$ is the number of GRBs in each sub-sample, $\rho$ is the Spearman correlation coefficient, $P$ is the probability that these correlations occur by chance, and $|\Delta\rho|$ is the absolute value of the change in percentage in the $\rho$ of each class relative to all GRBs. The variance is the population variance, defined as $\sigma^2=\frac{\sum(X-\mu^{2})}{N}$ for a given sub-sample of $N$ GRBs.}
%Gold+alpha
\label{bestfits}
\end{table}

The optical luminosity-time relation is defined as:

\begin{equation}
\log L_{\rm opt} = C_o + a_{\rm opt} \times \log T^{*}_{\rm opt},
\label{LumTime}
\end{equation}

\noindent where $C_{\rm o}$ is the normalization constant, and $a_{\rm opt}$ is the best-fit parameter representing the slope of the correlation in optical. To make the units dimensionless $T^{*}_{\rm opt}$ is divided by 1 s. The best-fit parameters of the total sample, and other sub-samples along with their squared scatter are shown in Table \ref{bestfits}. There are only four ULGRBs, so they are not included in Table \ref{bestfits}.
We also present in Table \ref{tab:parameters} the identity of the GRB, ID GRB, the redshift, $T_{90}$, the fitted parameters of the W07 model, the spectral index $\beta_{opt}$ and $\log L_{\rm opt}$ of the plateau phase.

\begin{deluxetable}{cccccccccc}
%\rotate
\tabletypesize{\scriptsize}
\tablecaption{Includes the identity of the GRB, ID GRB, its redshift, z, $T_{90}$, and the best-fit parameters calculated using the W07 model: the optical flux at the end of the plateau, $\log F_{opt}$, the end time of the plateau $\log T_{opt}$, the slope after the plateau, $\alpha_{opt}$, the optical spectral index, $\beta_{opt}$, and the optical luminosity at the end of the plateau, $L_{opt}$. The data source codes are first author followed by publication year: for example, Si2018 corresponds to \cite{Si2018}. Combined LCs have multiple authors listed. The full table with 102 GRBs is present at the following link: http://www.oa.uj.edu.pl/M.Dainotti/GRB2020/}
%TC:ignore
\tablehead{
\colhead{ID GRB}
&\colhead{$z$}
&\colhead{$T_{90}$}
&\colhead{class}
&\colhead{$\log F_{opt}$}
&\colhead{$\log T_{op}$}
&\colhead{$\alpha_{opt}$}
&\colhead{$\beta_{opt}$}
&\colhead{$\log L_{opt}$}
&\colhead{data source}
}
\startdata
%ID GRB & z & $T_{90}$ & class & $\log F_{opt}$ & $\log T_{opt}$ & $\alpha_{opt}$ & $\beta$ & $\log L_{opt}$ & data source \\ 
000301C & 2.03 & 2.00 & IS & -13.83 $\pm$ 0.12 & 5.88 $\pm$ 0.05 & 2.85 $\pm$ 0.14 & 0.59 $\pm$ 0.12 & 44.45 $\pm$ 0.14 & Si18 \\
000926 & 2.04 & 25.00 & L & -12.89 $\pm$ 0.03 & 5.12 $\pm$ 0.02 & 2.14 $\pm$ 0.04 & 1.01 $\pm$ 0.16 & 45.60 $\pm$ 0.08 & Kann06 \\
011211 & 2.14 & 270.00 & L & -13.52 $\pm$ 0.06 & 5.23 $\pm$ 0.04 & 1.98 $\pm$ 0.11 & 0.41 $\pm$ 0.14 & 44.72 $\pm$ 0.09 & Kann10 \\
021004 & 2.34 & 100.00 & L & -12.92 $\pm$ 0.02 & 5.40 $\pm$ 0.02 & 1.33 $\pm$ 0.03 & 0.67 $\pm$ 0.14 & 45.53 $\pm$ 0.08 & Li12, Li15 \\
030226 & 1.99 & 22.09 & L & -12.44 $\pm$ 0.04 & 3.24 $\pm$ 0.04 & 1.33 $\pm$ 0.05 & 0.57 $\pm$ 0.12 & 45.81 $\pm$ 0.07 & Kann06 \\
030328 & 1.52 & 199.20 & L & -12.70 $\pm$ 0.02 & 4.38 $\pm$ 0.02 & 1.25 $\pm$ 0.04 & 0.36 $\pm$ 0.45 & 45.22 $\pm$ 0.18 & Kann06 \\
030329 & 0.17 & 62.90 & SN-A & -11.76 $\pm$ 0.09 & 5.50 $\pm$ 0.05 & 1.46 $\pm$ 0.03 & 0.41 $\pm$ 0.17 & 44.11 $\pm$ 0.09 & Si18 \\
040924 & 0.86 & 2.39 & SN-C & -12.20 $\pm$ 0.04 & 3.50 $\pm$ 0.04 & 1.30 $\pm$ 0.02 & 0.63 $\pm$ 0.48 & 45.26 $\pm$ 0.13 & Kann06 \\
041006 & 0.72 & 17.40 & SN-C & -12.45 $\pm$ 0.03 & 4.08 $\pm$ 0.03 & 1.24 $\pm$ 0.01 & 0.36 $\pm$ 0.27 & 44.76 $\pm$ 0.07 & Si18 \\
050319 & 3.24 & 152.54 & XRR & -12.83 $\pm$ 0.02 & 4.44 $\pm$ 0.03 & 0.76 $\pm$ 0.03 & 0.76 $\pm$ 0.02 & 45.99 $\pm$ 0.02 & Zaninoni13 \\
050408 & 1.24 & 34.00 & L & -13.25 $\pm$ 0.03 & 4.36 $\pm$ 0.05 & 0.83 $\pm$ 0.05 & 0.28 $\pm$ 0.33 & 44.45 $\pm$ 0.12 & Si18 \\
050416A & 0.65 & 2.49 & XRF-D-IS-SN & -13.54 $\pm$ 0.05 & 4.15 $\pm$ 0.06 & 0.94 $\pm$ 0.08 & 0.92 $\pm$ 0.30 & 43.70 $\pm$ 0.08 & Li12, Li15 \\
050502A & 3.79 & 20.00 & L & -12.60 $\pm$ 0.04 & 3.72 $\pm$ 0.03 & 1.43 $\pm$ 0.02 & 0.76 $\pm$ 0.16 & 46.36 $\pm$ 0.11 & Kann10 \\
050525A & 0.61 & 8.83 & SN-B-XRR & -11.57 $\pm$ 0.04 & 3.90 $\pm$ 0.04 & 1.44 $\pm$ 0.03 & 0.52 $\pm$ 0.08 & 45.51 $\pm$ 0.04 & Kann10 \\
050603 & 2.82 & 21.00 & L & -11.88 $\pm$ 0.13 & 4.45 $\pm$ 0.08 & 1.85 $\pm$ 0.09 & 0.60 $\pm$ 0.00 & 46.71 $\pm$ 0.13 & Kann10 \\
050730 & 3.97 & 156.50 & L & -12.15 $\pm$ 0.04 & 4.34 $\pm$ 0.06 & 1.57 $\pm$ 0.07 & 0.52 $\pm$ 0.05 & 46.69 $\pm$ 0.05 & Kann10 \\
050801 & 1.56 & 19.40 & XRR & -10.98 $\pm$ 0.02 & 2.64 $\pm$ 0.02 & 1.19 $\pm$ 0.01 & 0.69 $\pm$ 0.34 & 47.09 $\pm$ 0.14 & Kann10 \\
050802 & 1.71 & 30.00 & L & -11.61 $\pm$ 0.08 & 2.91 $\pm$ 0.09 & 0.91 $\pm$ 0.01 & 0.36 $\pm$ 0.26 & 46.41 $\pm$ 0.14 & Kann10 \\
050820A & 2.61 & 244.69 & L & -11.97 $\pm$ 0.01 & 4.46 $\pm$ 0.02 & 1.02 $\pm$ 0.01 & 0.72 $\pm$ 0.03 & 46.62 $\pm$ 0.02 & Kann10; Zaninoni13 \\
050824 & 0.83 & 22.58 & XRF-E-SN & -12.50 $\pm$ 0.03 & 3.65 $\pm$ 0.06 & 0.65 $\pm$ 0.01 & 0.45 $\pm$ 0.18 & 44.87 $\pm$ 0.06 & Kann10 \\
050908 & 3.34 & 17.37 & XRR & -12.61 $\pm$ 0.08 & 3.26 $\pm$ 0.13 & 0.82 $\pm$ 0.08 & 1.25 $\pm$ 0.36 & 46.55 $\pm$ 0.24 & Zaninoni13 \\
050922C & 2.20 & 4.54 & IS & -11.65 $\pm$ 0.01 & 3.77 $\pm$ 0.01 & 1.25 $\pm$ 0.01 & 0.56 $\pm$ 0.01 & 46.69 $\pm$ 0.01 & Kann10; Zaninoni13; Oates09, Oates12 \\
051109A & 2.35 & 37.23 & L & -12.14 $\pm$ 0.03 & 3.74 $\pm$ 0.04 & 0.81 $\pm$ 0.02 & 1.06 $\pm$ 0.06 & 46.52 $\pm$ 0.04 & Zaninoni13 \\
051111 & 1.55 & 59.78 & L & -10.91 $\pm$ 0.03 & 2.77 $\pm$ 0.04 & 1.00 $\pm$ 0.04 & 0.76 $\pm$ 0.07 & 47.18 $\pm$ 0.04 & Si18 \\
060124 & 2.30 & 13.63 & XRR & -11.66 $\pm$ 0.03 & 3.63 $\pm$ 0.04 & 0.88 $\pm$ 0.00 & 0.75 $\pm$ 0.01 & 46.81 $\pm$ 0.03 & Zaninoni13 \\
060206 & 4.05 & 7.59 & XRR-IS & -12.05 $\pm$ 0.01 & 4.39 $\pm$ 0.01 & 1.39 $\pm$ 0.01 & 1.66 $\pm$ 0.05 & 47.62 $\pm$ 0.04 & Zaninoni13 \\
060210 & 3.91 & 255.00 & L & -11.70 $\pm$ 0.14 & 3.05 $\pm$ 0.08 & 1.49 $\pm$ 0.05 & 0.76 $\pm$ 0.00 & 47.30 $\pm$ 0.14 & Kann10 \\
060418 & 1.49 & 144.00 & XRR & -10.01 $\pm$ 0.09 & 2.35 $\pm$ 0.06 & 1.23 $\pm$ 0.01 & 0.69 $\pm$ 0.11 & 48.01 $\pm$ 0.10 & Kann10 \\
060512 & 0.44 & 8.49 & XRF & -12.44 $\pm$ 0.03 & 3.64 $\pm$ 0.05 & 0.74 $\pm$ 0.02 & 0.60 $\pm$ 0.00 & 44.35 $\pm$ 0.03 & Kann10 \\
060526 & 3.21 & 298.16 & XRR & -12.20 $\pm$ 0.01 & 4.19 $\pm$ 0.01 & 1.12 $\pm$ 0.01 & 0.65 $\pm$ 0.06 & 46.54 $\pm$ 0.04 & Kann10 \\
060605 & 3.78 & 114.79 & XRR & -11.16 $\pm$ 0.04 & 3.03 $\pm$ 0.05 & 1.04 $\pm$ 0.04 & 1.32 $\pm$ 0.03 & 48.18 $\pm$ 0.05 & Zaninoni13 \\
060607A & 3.07 & 99.30 & L & -11.78 $\pm$ 0.03 & 3.53 $\pm$ 0.04 & 1.25 $\pm$ 0.05 & 0.72 $\pm$ 0.27 & 46.97 $\pm$ 0.17 & Kann10 \\
060614 & 0.13 & 108.70 & KN-SEE-XRR & -13.05 $\pm$ 0.04 & 5.09 $\pm$ 0.02 & 2.15 $\pm$ 0.02 & 0.47 $\pm$ 0.04 & 42.53 $\pm$ 0.04 & Si18; Zaninoni13 \\
060714 & 2.71 & 114.99 & XRR & -12.47 $\pm$ 0.17 & 3.77 $\pm$ 0.21 & 0.76 $\pm$ 0.07 & 0.44 $\pm$ 0.04 & 45.99 $\pm$ 0.18 & Si18 \\
060729 & 0.54 & 115.35 & XRR-SN-E & -12.15 $\pm$ 0.03 & 5.07 $\pm$ 0.03 & 1.26 $\pm$ 0.06 & 0.85 $\pm$ 0.01 & 44.88 $\pm$ 0.03 & Zaninoni13 \\
060904B & 0.70 & 171.47 & XRR-SN-C & -12.11 $\pm$ 0.04 & 3.89 $\pm$ 0.04 & 1.20 $\pm$ 0.03 & 1.11 $\pm$ 0.10 & 45.25 $\pm$ 0.05 & Kann10 \\
060927 & 5.46 & 22.54 & XRR & -12.19 $\pm$ 0.26 & 3.24 $\pm$ 0.23 & 1.26 $\pm$ 0.06 & 0.82 $\pm$ 0.00 & 47.17 $\pm$ 0.26 & Kann10 \\
061007 & 1.26 & 75.31 & L & -8.76 $\pm$ 0.07 & 2.17 $\pm$ 0.03 & 1.75 $\pm$ 0.01 & 1.07 $\pm$ 0.19 & 49.23 $\pm$ 0.09 & Kann10 \\
061121 & 1.31 & 81.25 & L & -12.27 $\pm$ 0.04 & 3.85 $\pm$ 0.05 & 1.00 $\pm$ 0.01 & 0.68 $\pm$ 0.06 & 45.62 $\pm$ 0.05 & Zaninoni13 \\
070110 & 2.35 & 88.42 & XRR & -12.90 $\pm$ 0.06 & 4.44 $\pm$ 0.10 & 0.99 $\pm$ 0.05 & 0.60 $\pm$ 0.00 & 45.52 $\pm$ 0.06 & Kann10 \\
070125 & 1.55 & 60.00 & L & -12.25 $\pm$ 0.13 & 5.14 $\pm$ 0.03 & 2.37 $\pm$ 0.08 & 1.13 $\pm$ 0.02 & 45.99 $\pm$ 0.13 & Zaninoni13 \\
070208 & 1.17 & 64.00 & XRR & -12.37 $\pm$ 0.19 & 2.65 $\pm$ 0.32 & 0.52 $\pm$ 0.03 & 0.66 $\pm$ 0.00 & 45.40 $\pm$ 0.19 & Kann10 \\
070411 & 2.95 & 122.75 & XRR & -12.50 $\pm$ 0.19 & 3.38 $\pm$ 0.10 & 2.01 $\pm$ 0.31 & 1.17 $\pm$ 0.27 & 46.47 $\pm$ 0.25 & Zaninoni13 \\
070419A & 0.97 & 160.00 & XRF-SN-D & -12.67 $\pm$ 0.12 & 3.27 $\pm$ 0.07 & 1.40 $\pm$ 0.05 & 1.11 $\pm$ 0.22 & 45.05 $\pm$ 0.14 & Zaninoni13 \\
070810A & 2.17 & 11.03 & XRR & -12.45 $\pm$ 0.11 & 3.77 $\pm$ 0.12 & 1.50 $\pm$ 0.11 & 0.60 $\pm$ 0.00 & 45.90 $\pm$ 0.11 & Kann10 \\
071003 & 1.60 & 148.13 & L & -13.16 $\pm$ 0.07 & 5.51 $\pm$ 0.06 & 2.17 $\pm$ 0.15 & 0.35 $\pm$ 0.23 & 44.79 $\pm$ 0.12 & Kann10 \\
071010A & 0.99 & 6.20 & L & -11.12 $\pm$ 0.16 & 2.80 $\pm$ 0.17 & 0.81 $\pm$ 0.02 & 0.61 $\pm$ 0.12 & 46.47 $\pm$ 0.16 & Kann10 \\
071025 & 5.00 & 238.14 & XRR & -12.58 $\pm$ 0.03 & 3.37 $\pm$ 0.02 & 1.41 $\pm$ 0.01 & 0.93 $\pm$ 0.03 & 46.78 $\pm$ 0.03 & Kann10 \\
071031 & 2.69 & 180.89 & XRF & -11.99 $\pm$ 0.03 & 3.25 $\pm$ 0.03 & 0.85 $\pm$ 0.01 & 0.34 $\pm$ 0.30 & 46.41 $\pm$ 0.17 & Kann10 \\
071112C & 0.82 & 15.00 & SN-C & -11.73 $\pm$ 0.01 & 2.56 $\pm$ 0.02 & 0.92 $\pm$ 0.00 & 0.44 $\pm$ 0.11 & 45.64 $\pm$ 0.03 & Kann21b (in prep.) \\
%TC:endignore
\hline
\enddata\label{tab:parameters}
\vspace{3mm}
\end{deluxetable}

For the total sample, the resulting luminosity-time relation follows the form of Eq. \ref{LumTime} with constants:  $C_{\rm o} = 49.52\pm0.58$, $a_{\rm opt}=-1.02\pm0.16$, and $\sigma^{2} = 0.63$. The Spearman correlation coefficient, $\rho=-0.77$, and the probability of this correlation occurring by chance, P, is $2.7\times10^{-23}$.
For all classes $\rho$ is very high and $P \ll 0.05$. This behavior is consistent across all classes, thus guaranteeing that this correlation holds regardless of class.
The luminosity-time correlation holds in optical afterglows even for this sample of 102 GRBs, which is the largest compilation of optical plateaus so far in the literature. The slopes of the luminosity-time correlation in X-ray and optical for a common overlapping sample agree within $1\sigma$, $a_{\rm X}=-1.32\pm0.28$ and $a_{\rm opt}=-1.12\pm0.26$, thus we can infer that the energy reservoir of the GRB during the plateau in both electromagnetic regimes is constant and is independent of class (the best-fit slopes through each of the classes are $a\approx-1$; see Table \ref{bestfits}).

The Gold Sample has a $\sigma^{2}=0.30$, smaller than that of the total sample by  $52.4\%$. To compare the tightness of the correlation in optical and in X-rays, we identify the GRBs coincident between our optical sample and the X-ray sample of \cite{Srinivasaragavan2020} and
\cite{Dainotti2020arXiv}; the two samples have 56 GRBs in common. From the fit of these 56 GRBs we obtain the following X-ray and optical parameters: $C_{0,X}= 52.02 \pm 0.99$, $a_{X}=1.32 \pm 0.28$, while $C_{0,opt}= 49.91 \pm 0.91$, $a_{opt}=1.12 \pm 0.26$. This leads us to conclude that the luminosity-time correlation in X-rays is tighter than in optical. Since in both cases within errors the slope of the correlation is compatible with $-1$, this implies that the energy reservoir of the plateau is constant and that a magnetar scenario can be the leading explanation for the optical correlation as well as for the X-ray one.

The first panel of Figure \ref{classanglet90} shows our sample divided by class. No class clusters in a particular region of the plot. Indeed, both the slope $a_{\rm opt}$ and the normalization agree within $1\sigma$ for all classes; $\rho$ for all classes are shown in Table \ref{bestfits}. The gold class has the highest correlation coefficient and the smallest squared scatter, $\sigma^{2}=0.30$, with a percentage decrease compared to all GRBs of $52.4\%$, see last column of Table 1. This is aligned with a previous result shown in \cite{Dainotti2017b, Dainotti2016}: the gold sample has a much higher correlation coefficient, and a smaller scatter also in X-rays. 

The second panel of Fig. \ref{classanglet90} shows the distinction between Type I and Type II GRBs. 

The third panel of Fig. \ref{classanglet90} represents all GRBs binned by the angle of inclination of the plateau feature. For each of the angle bins in increasing order $\rho=(-0.65, -0.78, -0.86, -0.83)$, where the third bin ($38.7^{\circ} < \theta \leq 48.6^{\circ}$, black triangles in figure) shows the tightest correlation.

\begin{figure*}[!t]
\centering
\includegraphics[width=0.6\textwidth,angle=0,clip]{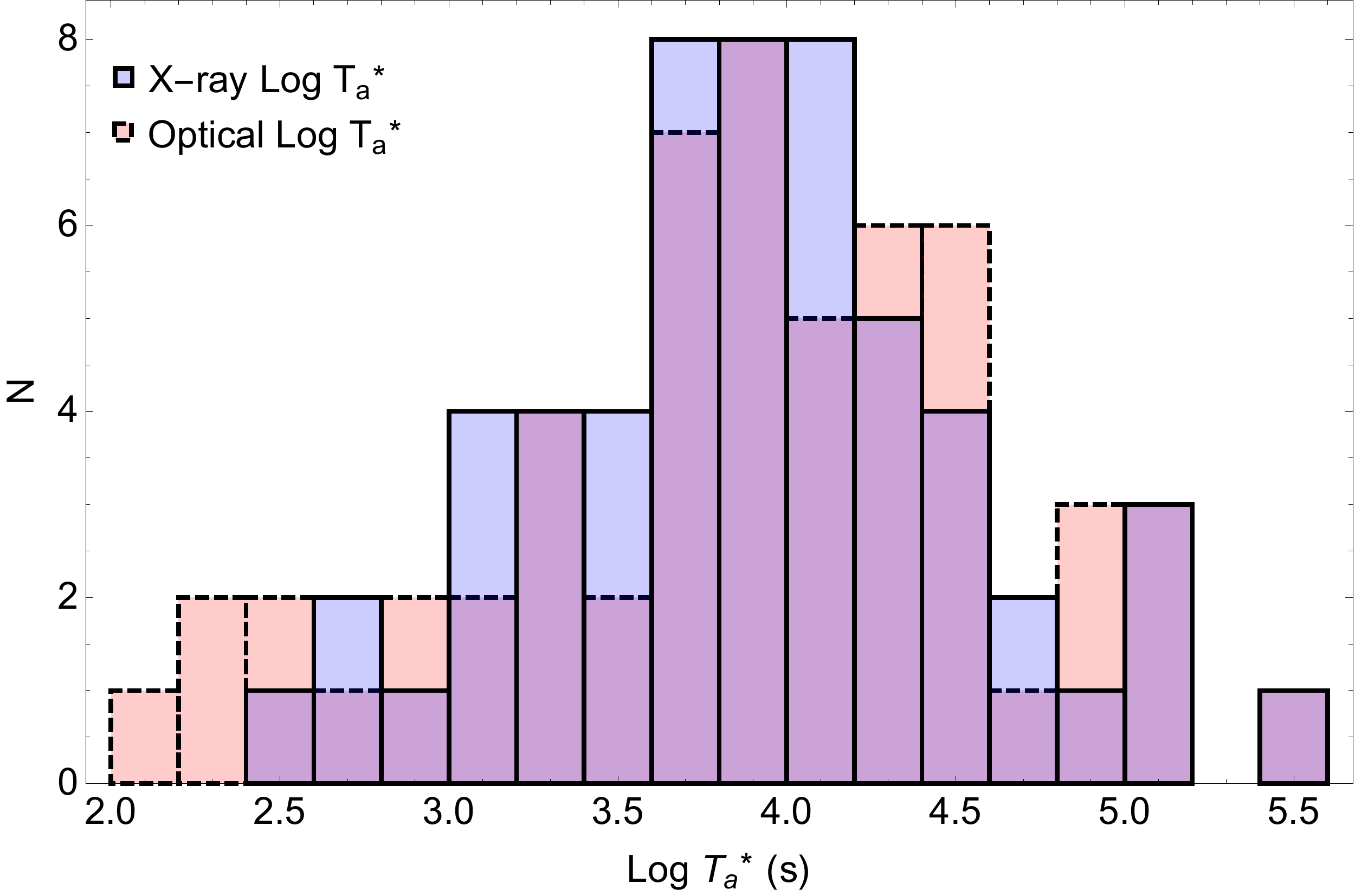}
\caption{The differential histogram of the end-time of the plateau in the rest-frame for the optical (in red) and X-ray (in blue) afterglow. The violet colours are the ones coincident between X-rays and optical.}
\label{histogram}
\end{figure*}
\noindent 

The fourth panel of Figure \ref{classanglet90} shows all GRBs divided by $T^{*}_{90}$; $\rho$ for each of the $T^{*}_{90}$ bins, in increasing order, are $\rho=(-0.73, -0.66, -0.85, -0.75)$. The third bin ($62.9s < T^{*}_{90} < 148.0s$) has the highest monotonic correlation.
%{\bf Since the segregation in classes does not constrain further the scatter of the correlation we plan in a forthcoming paper to follow the same approach in \cite{Dainotti2016, Dainotti2017a, Dainotti2020arXiv} and to extend this correlation in 3D with the peak prompt optical luminosity whenever is possible to attempt a considerable reduction of the scatter in the correlation as it occurred in X-rays}.

\section{\label{conclusion} Discussion and Conclusions}

We have gathered the largest compilation of optical plateaus to date (102 GRBs) and shown that the $L_{\rm opt}-T^{*}_{\rm opt}$ correlation holds for a sample which is more than double the largest sample presented in the literature. The optical correlation is:
\begin{equation}
\log L_{\rm opt} = (49.52\pm0.58) -(1.02 \pm 0.16)\times \log T^{*}_{opt}
\end{equation}
\noindent with $\sigma^{2} = 0.63$ and $\rho=-0.77$ for the whole sample. The Gold Sample has a reduced $\sigma^{2} = 0.30$ of $52.4\%$ and an increased $\rho=-0.86$ ($10.5\%$ increase, see Table 1 for the absolute value of $\Delta_{\rho}$). The slopes of the X-ray and optical luminosity-time correlation are within 1$\sigma$; both demonstrate strong linear anti-correlations. Given the slope of the correlation is nearly $-1$, this further supports that the plateau has a fixed energy reservoir independent of a given class and a possible explanation can be the magnetar model. 
The source of the scatter of the correlation comes both from a physical point of view, depending on the energy mechanism underlying the plateau, which regime and frequency, and from an instrumental point of view. We indeed obtain a reduced scatter when we consider LCs belonging to the Gold Sample. Additionally, we find that the $L_{\rm opt}-T^{*}_{\rm opt}$ correlation holds regardless of GRB class, plateau angle, or $T^{*}_{90}$.

Furthermore, we find that the end-time of the plateau is achromatic between X-ray and optical observations for a sub-sample of GRBs observed in both bands (see Fig. \ref{histogram}). It is compelling that the candidate feature, the plateau, to standardize GRBs is achromatic between the X-rays and optical, the two wavelengths in which the majority of plateaus are observed. This analysis can be ascribed to a larger context for the determination of whether or not the plateau is achromatic, since some cases of plateaus have been also observed by the {\it Fermi}-LAT in high-energy gamma-rays (\citealt{Ajello2019}).

\acknowledgements
This work made use of data supplied by the UK \emph{Swift} Science Data Centre at the University of Leicester. We thank G. Sarracino for his help in modifying our host extinction code in Python, R. Wagner for the fitting of some GRB LCs, L. and A. Zambrano Tapia, M. Fuentes Qui\~{n}onez, and E. Fern\'andez Guzm\'ann for the help in bibliography and combining some LCs. M.G.D acknowledges the American Astronomical Society Chretienne Grant for its initial support. D.A.K. acknowledges support from Spanish National Research Project RTI2018-098104-J-I00 (GRBPhot). We thank E. Cuellar for his guidance and his work in organizing the SULI summer program. This work was supported in part by the U.S. Department of Energy, Office of Science, Office of Workforce Development for Teachers and Scientists (WDTS) under the Science Undergraduate Laboratory Internships (SULI) program. Parts of this research were conducted by the Australian Research Council Centre of Excellence for Gravitational Wave Discovery (OzGrav), through project number CE170100004.

\bibliographystyle{yahapj}
\bibliography{references}
%\nocite{*}

%\input{referencesold.tex}

% \bibliographystyle{aasjournal}
% \bibliography{bib_Swiftpaper.tex.bib}

% \begin{thebibliography}

% \bibliography{REFERENCES.tex}
% \bibliographystyle{yahapj}

% this works!!!
% \printbibliography
% \bibliographystyle{yahapj}
% \nocite{*}

\end{document}